\documentclass{cidr-2020}

\usepackage{amsmath}
\usepackage{amssymb}

\usepackage{graphicx}
\usepackage{subfigure}
\usepackage[np,autolanguage]{numprint}
\usepackage{units}
\usepackage{balance}
\usepackage{algorithm}
\usepackage[noend]{algorithmic}

\usepackage[pdftex,pdfpagelabels=false]{hyperref} 
\hypersetup{
    unicode=false,          
    pdftoolbar=true,        
    pdfmenubar=true,        
    pdffitwindow=true,      
    pdfstartview={FitV},    
    pdftitle={SystemDS: A Declarative Machine Learning System for the End-to-End Data Science Lifecycle},    
    pdfauthor={Matthias Boehm, Iulian Antonov, Sebastian Baunsgaard, Mark Dokter, Robert Ginth{\"o}r, Kevin Innerebner, Florijan Klezin, Stefanie Lindstaedt, Arnab Phani, Benjamin Rath, Berthold Reinwald, Shafaq Siddiqi, Sebastian Benjamin Wrede}, 
    pdfsubject={},   
    pdfcreator={},   
    pdfproducer={}, 
    pdfkeywords={}, 
    pdfnewwindow=true,      
    bookmarksnumbered=true, 
    bookmarksopen=true,     
    bookmarksopenlevel=2,   
    colorlinks=false,        
    linkcolor=red,        
    citecolor=green,         
    filecolor=black,        
    urlcolor=black          
}

\newcommand{\mat}[1]{\ensuremath{\mathbf{#1}}}

\newcommand{\mb}{\unit{\,MB}}
\newcommand{\gb}{\unit{\,GB}}

\newcommand{\gflop}{\unit{\,GFLOP}} 

\newenvironment{itemize2}{\begin{itemize}\setlength{\itemsep}{2.5pt}\setlength{\parskip}{0pt}\setlength{\parsep}{0pt}}{\end{itemize}}

\renewcommand{\footnotesize}{\small}
\newtheorem{example}{Example}

\begin{document}



\title{SystemDS: A Declarative Machine Learning System\\ for the End-to-End Data Science Lifecycle}

\numberofauthors{1} 
\author{
\alignauthor Matthias Boehm\textsuperscript{1,2}, Iulian Antonov\textsuperscript{2}, Sebastian Baunsgaard\textsuperscript{1}\thanks{Work done while at IT University of Copenhagen, Denmark.}, Mark Dokter\textsuperscript{2}, Robert Ginth{\"o}r\textsuperscript{2}, Kevin Innerebner\textsuperscript{1}, Florijan Klezin\textsuperscript{2}, Stefanie Lindstaedt\textsuperscript{1,2}, Arnab Phani\textsuperscript{1}, Benjamin Rath\textsuperscript{1}, Berthold Reinwald\textsuperscript{3}, Shafaq Siddiqi\textsuperscript{1}, Sebastian Benjamin Wrede\textsuperscript{2}\footnotemark[1]\\~\vspace{-0.15cm}\\ 
\affaddr{\textsuperscript{1} Graz University of Technology;~~Graz, Austria}\\ 
\affaddr{\textsuperscript{2} Know-Center GmbH;~~Graz, Austria}\\
\affaddr{\textsuperscript{3} IBM Research -- Almaden;~~San Jose, CA, USA}\\
}

\maketitle
\begin{abstract}
Machine learning (ML) applications become increasingly common in many domains. ML systems to execute these workloads include numerical computing frameworks and libraries, ML algorithm libraries, and specialized systems for deep neural networks and distributed ML. These systems focus primarily on efficient model training and scoring. However, the data science process is exploratory, and deals with underspecified objectives and a wide variety of heterogeneous data sources. 
Therefore, additional tools are employed for data engineering and debugging, which requires boundary crossing, unnecessary manual effort, and lacks optimization across the lifecycle.
In this paper, we introduce SystemDS, an open source ML system for the end-to-end data science lifecycle from data integration, cleaning, and preparation, over local, distributed, and federated ML model training, to debugging and serving. To this end, we aim to provide a stack of declarative language abstractions for the different lifecycle tasks, and users with different expertise. We describe the overall system architecture, explain major design decisions (motivated by lessons learned from Apache SystemML), and discuss key features and research directions. 
Finally, we provide preliminary results that show the potential of end-to-end lifecycle optimization.
\end{abstract}

\section{Introduction}

Machine learning (ML) applications profoundly transform our lives, and many domains such as health care, finance, media, transportation, production, and information technology itself. 
Increased digitalization, sensor-equipped vehicles and production pipelines, feedback loops in data-driven products, and data augmentation also provide large, labeled data collections for training the underlying ML models. 

\textbf{Existing ML Systems:} ML systems to execute these workloads are---due to a variety of ML algorithms and lack of standards---still diverse and rapidly evolving. Major system categories include numerical computing frameworks like R, Python NumPy~\cite{WaltCV11}, or Julia~\cite{BezansonEKS17}, algorithm libraries like Scikit-learn~\cite{PedregosaVGMTGBPWDVPCBPD11} or Spark MLlib~\cite{MengBYSVLFTAOXX16}, large-scale linear algebra systems like Apache SystemML~\cite{BoehmDEEMPRRSST16} or Mahout Samsara~\cite{MahoutSamsara}, and specialized deep neural network (DNN) frameworks like TensorFlow~\cite{AbadiBCCDDDGIIK16}, MXNext~\cite{ChenLLLWWXXZZ15}, or PyTorch~\cite{pytorch,PyTorch2}. These systems primarily rely on numeric matrices or tensors, and focus on efficient ML training and scoring.

\textbf{Exploratory Data-Science Lifecycle:} In contrast to classical ML problems, the typical data science process is exploratory. Stakeholders pose open-ended problems with underspecified objectives that allow different analytics, and can leverage a wide variety of heterogeneous data sources \cite{PolyzotisRWZ18}. Data scientists then investigate hypotheses, integrate the necessary data, run different analytics, and look for interesting patterns or models \cite{CohenDDHW09}. Since the added value is unknown in advance, little investment is made into systematic data acquisition, and preparation. This lack of infrastructure results in redundancy of manual efforts and computation, especially in small or medium-sized enterprises, which often lack curated catalogs of data and artifacts.

\textbf{Data Preparation Problem:} It is widely recognized that data scientists spend 80-90\% of their time finding relevant datasets, and performing data integration, cleaning, and preparation tasks \cite{StonebrakerI18}. For this reason, many industrial-strength ML applications have dedicated subsystems for data collection, verification, and feature extraction \cite{BaylorBCFFHHIJK17,SchelterLSCBG18,SculleyHGDPECYC15}. Since data integration and cleaning are, however, stubbornly difficult tasks to automate \cite{BernsteinM07}, existing work primarily focuses on well-defined subproblems or---like Wrangler \cite{KandelPHH11,RamanH01} and Trifacta~\cite{HeerHK15}---on semi-manual data wrangling through interactive UIs. Unfortunately, this diversity of tools and specialized algorithms lacks broad systems support, requires boundary crossing, and lacks optimization across the lifecycle. These problems motivated various in-database ML toolkits \cite{ChengQR12,FengKRR12,HellersteinRSWFGNWFLK12,LuoGGPJ17,PassingTHLSGK017,DsilvaMK18} to enable data preparation and ML training/scoring in SQL. However, this approach was---except for success stories like factorized learning \cite{KumarNP15,NikolicO18,SchleichOC16}---mostly unsuccessful because data scientists perceived in-database ML and array databases \cite{StonebrakerBPR11} as unnatural and cumbersome due to the need for data loading, and the verbosity of linear algebra in SQL \cite{DsilvaMK18}.

\textbf{A Case for Declarative Data Science:} From the viewpoint of a data scientist, it seems most natural to specify data science lifecycle tasks through familiar R or Python syntax and use stateless systems, which directly process files or in-memory objects. A key observation is that state-of-the-art data integration algorithms (e.g., for data extraction, schema alignment, entity linking, and data fusion) are themselves based on machine learning \cite{DongR18}. Similar observations can be made for data cleaning \cite{HeidariMIR19,RekatsinasCIR17}, outlier detection \cite{ChandolaBK09,HodgeA04}, missing value imputation \cite{CambroneroFSM17,mice}, semantic type detection \cite{HulsebosHBZSKDH19,abs-1911-06311}, data augmentation \cite{autoaugment}, feature selection \cite{VenablesR02}, model selection and hyper-parameter tuning (e.g., via Bayesian Optimization) \cite{FeurerKES0H19, KotthoffTHHL17, OlsonM19}, and model debugging \cite{ChungKPTW19,GebalyAGKS14}. We aim to leverage this characteristic by extending ML systems with high-level abstractions for the entire data science lifecycle but implement these abstractions with a domain-specific language (DSL) used for ML training and scoring. As a ``byproduct'', we avoid boundary crossing and the system can perform optimizations across lifecycle tasks.

\textbf{Contributions:} Following this goal of better systems support for declarative data science pipelines, we introduce SystemDS\footnote{The source code and releases (SystemDS 0.1 published 08/2019) are available at \url{https://github.com/tugraz-isds/systemds}.}, an open source ML system for the end-to-end data science lifecycle from data integration, cleaning, and preparation, over efficient ML model training, to debugging and serving. Our detailed contributions are:
\begin{itemize2}
\item \emph{Lessons Learned and Vision:} We first reflect on lessons learned from building Apache SystemML (as the predecessor of SystemDS), open problems, and how they influenced the overall vision of SystemDS in Section~\ref{sec:vision}.
\item \emph{System Architecture:} Following the outlined vision, we then describe the resulting system architecture and design decisions regarding language abstractions, compilation and runtime backends, as well as the underlying data model of heterogeneous tensors in Section~\ref{sec:sysarch}.
\item \emph{Key Features and Directions:} Subsequently, we discuss key features like lineage tracing, data preparation primitives, and federated ML in Section~\ref{sec:dirs}.
\item \emph{Preliminary Results:} Finally, we present preliminary results comparing performance with TensorFlow and Julia, and showing optimization opportunities---such as reuse---across lifecycle tasks in Section \ref{sec:exp}.
\end{itemize2}

\section{Lessons Learned and Vision}
\label{sec:vision}

Our central goal is to provide high-level abstractions and dedicated system support for the entire data science lifecycle, with a special focus on ML pipelines. Existing end-to-end ML frameworks like TFX \cite{BaylorBCFFHHIJK17}, KeystoneML \cite{SparksVKFR17}, or Alpine Meadow \cite{ShangZBKECBUK19} are built on top of ML libraries, which allows reusing these evolving systems, but consequently view ML algorithms as black boxes. Cross-library compilation in Weld \cite{PalkarTSSAZ17} focuses primarily on UDFs. In contrast, we believe that control of the compiler and runtime is of utmost importance for seamless interoperability and performing optimizations such as fine-grained redundancy elimination. For this reason, we forked SystemDS from Apache SystemML \cite{BoehmBERRSTT14,BoehmDEEMPRRSST16,GhotingKPRSTTV11} and we are currently rebuilding its foundations.

\subsection{Lessons Learned from SystemML}

SystemML has been under active development---with fluctuating team size---for about a decade. 
Here, we share selected lessons learned that influenced the vision, design, and system architecture of SystemDS:
\begin{itemize2}
\item \emph{L1 Data Independence \& Logical Operations:} Physical data independence and high-level linear algebra operations provided great independence of the evolving technology stack (e.g., MR$\rightarrow$Spark, and GPUs), simplified development (e.g., library algorithms) and deployment (e.g., large-scale/embedded), and enabled adaptation to changing cluster and data characteristics (e.g., local/distributed, and dense/sparse/compressed).
\item \emph{L2 User Categories:} SystemML focused on linear algebra programs for algorithm developers and ML researchers who write new or customize existing ML algorithms. However, this area is a niche as most data scientists work with existing algorithms, but need better support for other lifecycle tasks instead.
\item \emph{L3 Diversity of ML Algorithms \& Apps:} Today's ML systems literature largely focuses on DNNs, mini-batch SGD, and parameter servers. In practice, however, there is a wide variety of existing ML algorithms (e.g., unsupervised, (semi-)supervised batch 1\textsuperscript{st}/2\textsuperscript{nd}-order, ensembles, mini-batch DNNs, hybrid batch)---which require very different parallelization strategies---as well as complex ML applications that combine ML algorithms, numerical computing, and rules.
\item \emph{L4 Heterogeneous \& Structured Data:} SystemML supports feature transformations on frames (2D-tables with a schema). However, many applications deal with heterogeneous data (e.g., multi-modal), various forms of structure, and a wide variety of data corruptions. Boundary crossing for integrating, and preparing these datasets is still a major issue.
\end{itemize2}

\textbf{Discussion:} A natural question is why SystemML ultimately did not---except for few IBM products and services---reach adoption in practice. There are a number of overlapping reasons. First of all, the focus on ML researchers who directly experiment with large data was a niche (\emph{L2}). Organizations that deal with large, distributed datasets often have dedicated teams or use existing libraries. Over the last years, the ML research focus also moved toward mini-batch DNN workloads, parameter servers, and almost exclusively Python bindings (\emph{L3}). Together, these developments rendered SystemML's key differentiator---of automatically compiling R-like scripts into hybrid runtime plans of local and distributed Spark operations---ineffective in spurring adoption. Lacking a pressing need, users gravitated toward more popular frameworks like Scikit-learn, Spark MLlib, PyTorch, and TensorFlow. Although SystemML's R-like (and later Python-like) syntax was chosen to simplify adoption, it was still a DSL, with all its challenges like limited documentation and online resources, as well as difficulties of building an optimizing compiler, especially for unknown workloads. 

\subsection{SystemDS Vision and Design}
  
\textbf{Vision:} In contrast to traditional ML training and scoring, there is a dire need for more effective data integration, cleaning, and preparation as well as model debugging, especially for large-scale problems. Accordingly, the central goal of SystemDS is to provide high-level abstractions and systems support for the wealth of data science lifecycle tasks (\emph{L3} and \emph{L4}) and users with different expertise (\emph{L2}). Our overall vision comprises three components. First, we aim to implement a hierarchy of abstractions for data science tasks based on a \emph{DSL for ML training and scoring} (Section~\ref{sec:apis}) because state-of-the-art data integration, preparation, and cleaning heavily rely on machine learning; because exploratory data science interleaves data preparation, ML training, scoring, and debugging in an iterative process; and because once these tasks are expressed in dense or sparse linear algebra, we expect very good performance. Second, we aim to provide a holistic system infrastructure for the different lifecycle tasks and algorithm classes that require different parallelization strategies. The key to accomplish that are \emph{complementary runtime backends and an optimizing compiler} (Section~\ref{sec:runtime}). We take advantage of the now relatively mature SystemML compiler and extend it for new architectures like federated learning. Furthermore, the hierarchy of language abstractions inevitably creates fine-grained redundancy, which we aim to eliminate via automatic optimization at compiler and runtime level. Third, supporting data integration and preparation in linear algebra programs requires a more general data model for handling heterogeneous and structured data. In contrast to existing ML systems, our central data model are \emph{heterogeneous tensors} (Section~\ref{sec:data}), i.e., multi-dimensional arrays of different data types, including JSON strings to represent nested data.

\textbf{Summary:} Together, we believe that a holistic system infrastructure for effective and efficient data preparation, ML training and debugging (e.g., distributed data cleaning under awareness of the entire pipeline)---something that cannot be composed from existing libraries---addresses a pressing issue.
At the same time, there are lots of open challenges that require novel techniques throughout the system stack.

\section{System Architecture} 
\label{sec:sysarch}

The outlined vision directly influenced the design and architecture of SystemDS. We now provide an overview of language abstractions, runtime backends, and the underlying data model, before describing selected key features. 

\subsection{Language Abstractions and APIs}
\label{sec:apis}

\begin{figure}[!t]
  \centering
	\includegraphics[scale=0.445]{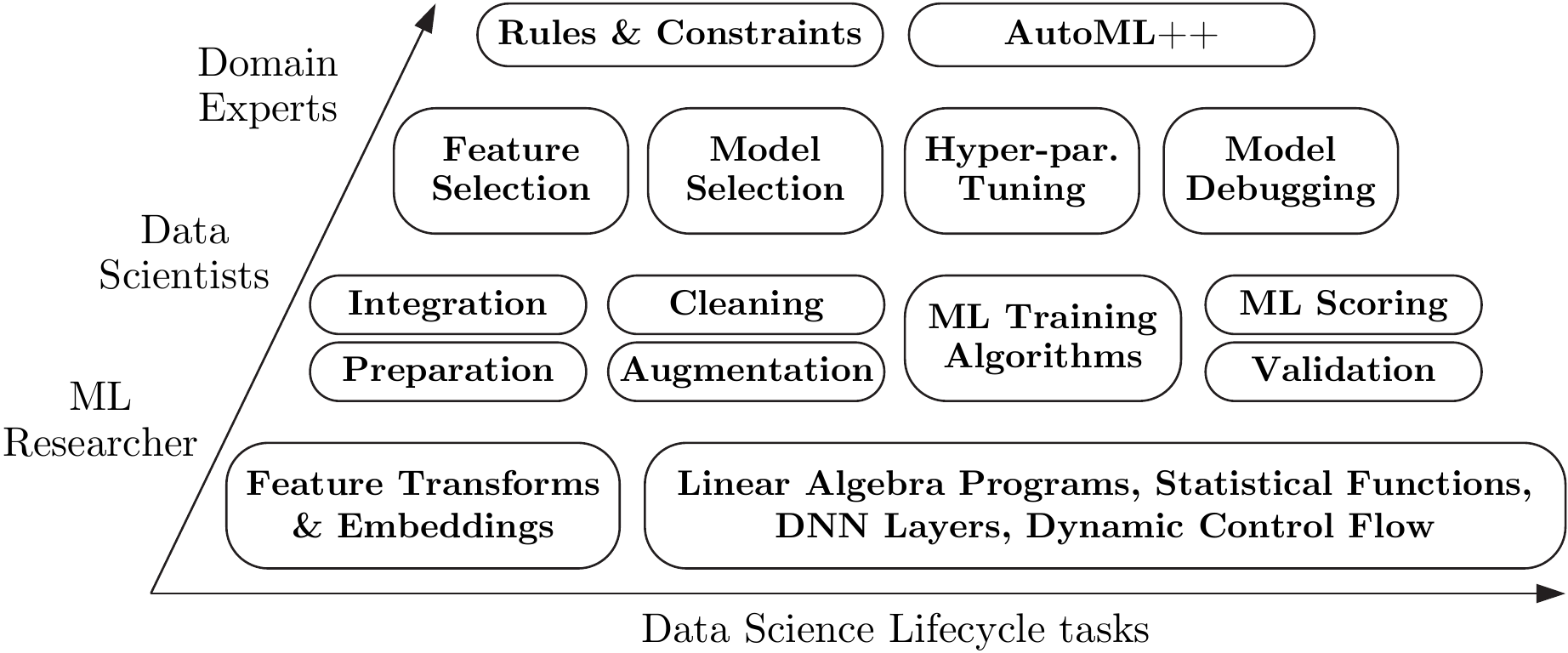}
	\vspace{-0.2cm}
	\caption{\label{fig:stack}A Stack of Declarative Languages.}
\end{figure}

\textbf{Scripting Language:} Following the success of declarative ML (\emph{L1}), we leverage SystemML's DML (Declarative ML Language) \cite{GhotingKPRSTTV11}, a scripting language with R-like syntax for linear algebra, aggregations, element-wise and statistical operations, control flow programs, and user-defined functions. However, we extend this language---as shown in Figure~\ref{fig:stack}---by a stack of declarative abstractions for different lifecycle tasks, and users with different expertise (\emph{L2}). We aim to provide data scientists and domain experts with abstractions for data integration and extraction, cleaning and preparation, data augmentation, model validation, model selection, hyper-parameter tuning, model debugging, rules and AutoML with domain-specific extensions (e.g., constraints and simulation models). To facilitate the development and compilation of these abstractions, we introduced a mechanism for registering DML-bodied built-in functions, and we aim to advance existing size propagation techniques.

\begin{figure}[!t]
  \centering
	\includegraphics[scale=0.443]{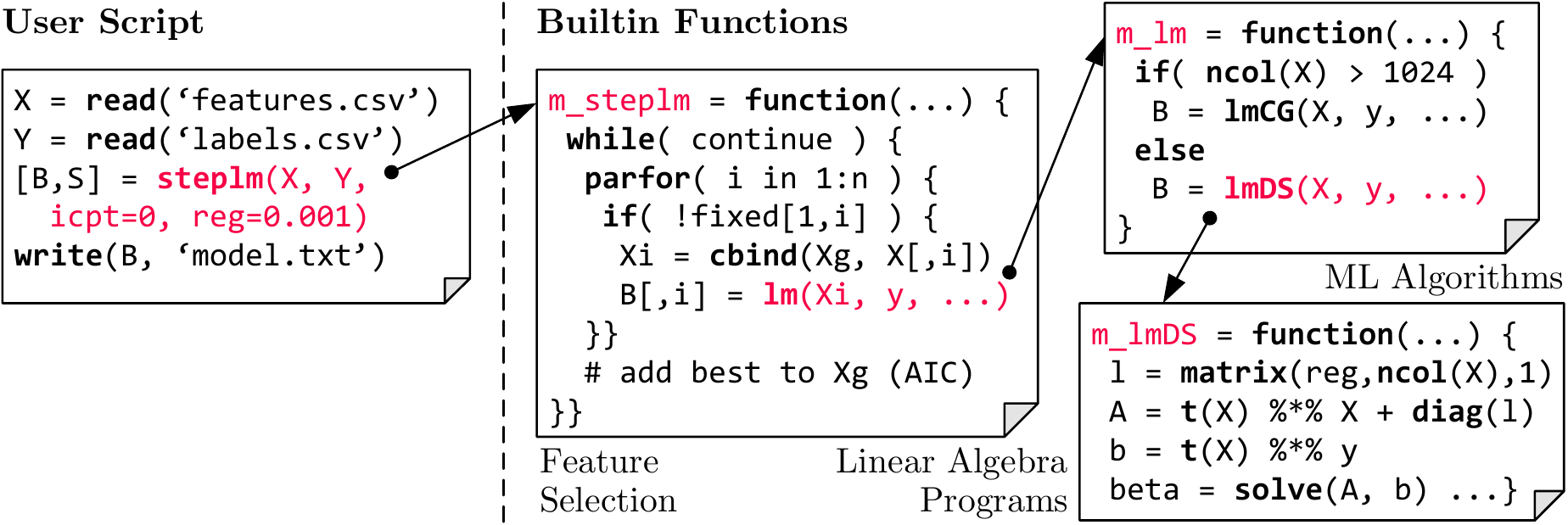}
	\vspace{-0.2cm}
	\caption{\label{fig:steplm}Example Stepwise Linear Regression.}
\end{figure}

\begin{example}[Stepwise Linear Regression] \label{ex:1} To see how powerful these abstractions are, consider stepwise linear regression \cite{VenablesR02}, a classical forward feature selection method. This method iteratively runs what-if scenarios and greedily selects the next best feature until the Akaike information criterion (AIC) does not improve anymore (see \texttt{steplm} in Figure~\ref{fig:steplm}). Each configuration trains a regression model via \texttt{lm}, which in turn calls iterative or closed form linear algebra programs. For an input matrix $\mat{X}$ (e.g., with $\text{ncol}(\mat{X})=500$ features), the compiler then collapses these abstractions---by removing unnecessary branches, dead code elimination, and function inlining---compiles distributed operations if necessary, and can reason about the end-to-end computation.
\end{example}

\textbf{APIs and Language Bindings} (Fig.~\ref{fig:arch}-1)\textbf{:} The user-defined scripts can then be executed with different APIs as shown in Figure~\ref{fig:arch}, where gray-shaded boxes indicate major new components. This includes command line invocation (e.g., through \texttt{java} or \texttt{spark-submit}) and the programmatic APIs (\texttt{MLContext} or \texttt{JMLC}). The \texttt{MLContext} API allows Spark RDDs and Datasets as script inputs, while \texttt{JMLC} is an API for embedded, low-latency scoring that allows pre-compiling a script and repeatedly executing it with different in-memory inputs. For a seamless integration with typical data science workflows, we will further add host language bindings for Python, R, and Java. These bindings expose individual operations, internally collect larger DAGs of operations and entire programs, and finally compile and execute efficient runtime plans on user request or output conversion.

\subsection{Compiler and Runtime Operations}
\label{sec:runtime}

\begin{figure}[!t]
  \centering
	\includegraphics[scale=0.443]{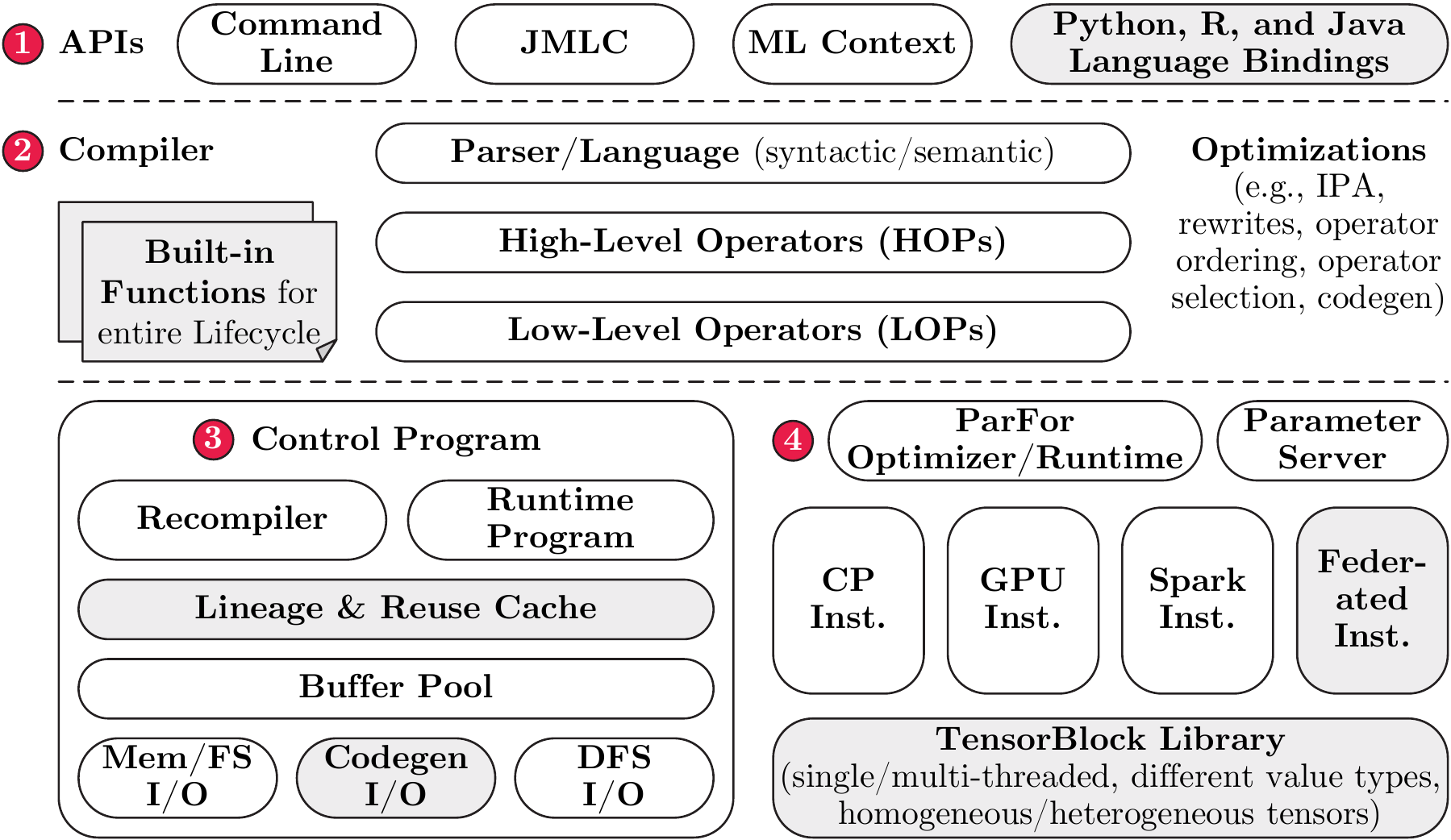}
	\vspace{-0.2cm}
	\caption{\label{fig:arch}SystemDS Architecture and Components.}
\end{figure}

\textbf{Compilation Chain} (Fig.~\ref{fig:arch}-2)\textbf{:} SystemDS inherits SystemML's compilation chain \cite{BoehmBERRSTT14}. Each user script or DML-bodied function is compiled into a hierarchy of statement blocks and statements, where control flow statements like loops or branches delineate these blocks. All statements of a basic (i.e., last-level) block are compiled into a DAG of high-level operators (HOPs), which represent logical operations. After multiple rounds of rewrites, size propagation (of dimension and sparsity), and operator ordering, we then compute memory estimates for each operation. Based on these  estimates, we in turn decide for local or distributed operations, and construct a DAG of low-level---i.e., physical---operators (LOPs). Finally, we create an executable program of program blocks and a sequence of runtime instructions---similar to MAL plans in MonetDB \cite{IvanovaKNG09}---per block.

\textbf{Runtime Control Program} (Fig.~\ref{fig:arch}-3)\textbf{:} The compiled runtime program is interpreted as a so-called control program (CP) in the client or Spark driver process. Besides program block and instruction execution, the control program also performs dynamic recompilation (recompilation of basic blocks to mitigate initial unknowns similar to adaptive query processing \cite{DeshpandeIR07}), and maintains a multi-level buffer pool that is responsible for evicting intermediate variables if necessary, persistent reads and writes from and to distributed file systems like HDFS or S3, and data exchange between the different runtime backends. Major CP extensions are built-in support for data provenance and generated I/O primitives for external formats as discussed in Sections~\ref{sec:lineage} and \ref{sec:preparation}.
 
\textbf{Runtime Operations} (Fig.~\ref{fig:arch}-4)\textbf{:} With the diversity of ML algorithms and apps (\emph{L3}) in mind, we further extend SystemML's multiple backends. We include local CPU and GPU instructions, as well as distributed Spark instructions. In addition, we introduce a new class of federated instructions as discussed in Section~\ref{sec:federated}. These instructions rely on a common \texttt{TensorBlock} operation library, which extends SystemDS from numeric matrices to heterogeneous, multi-dimensional arrays as described in Section~\ref{sec:data}. For local operations, such a block holds the entire tensor, while distributed tensors are represented as RDD collections of fixed-sized blocks. Besides reuse, this approach also ensures consistency across local and distributed operations. Additionally, we support dedicated backends for parallel for loops \cite{BoehmTRSTBV14} (e.g., for hyper-parameter tuning, and cross validation), and parameter servers (e.g., for mini-batch DNN training). The changed data representation necessitates major changes throughout the entire compiler and runtime stack.

\subsection{Data Model: Heterogeneous Tensors}
\label{sec:data}

\textbf{Data Model Motivation:} Our goal of supporting the end-to-end data science lifecycle poses two main requirements on the underlying data model. First, we need to represent heterogeneous and structured datasets for data integration and preparation (\emph{L4}). Second, many lifecycle tasks and ML algorithms benefit from native support of multi-dimensional arrays. Therefore, and in contrast to most existing DNN frameworks and array databases---which support homogeneous arrays (e.g., tensors of floats or integers)---our data model is a heterogeneous tensor, that is, a multi-dimensional array where one dimension has a schema. We believe this is more natural than 2D datasets because it allows for range indexing that guarantees matching dimensions for subsequent operations (e.g., matrix multiplication).

\begin{figure}[!t]
	\centering
	\hfill
	\subfigure[Heterogeneous TensorBlock]{
	   \label{fig4a}\includegraphics[scale=0.505]{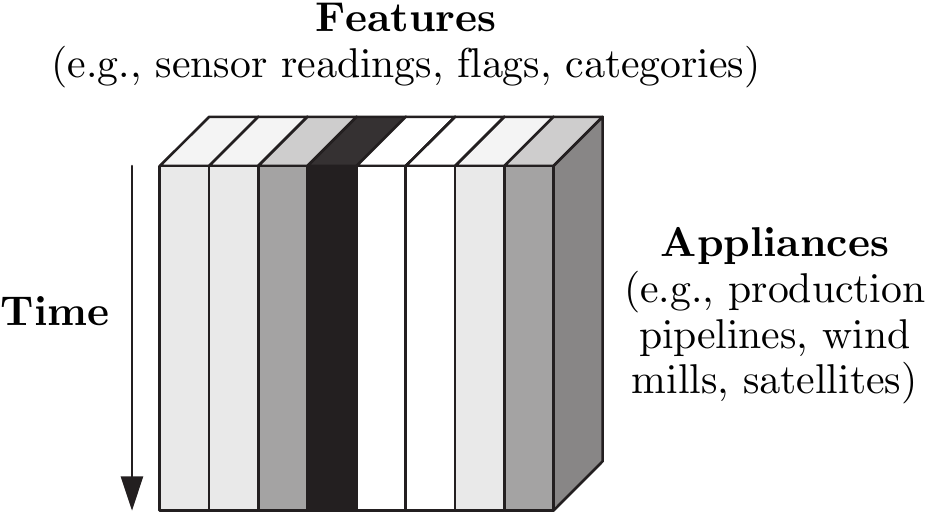}}
	\hfill	
	\subfigure[Distributed Tensor]{
	   \label{fig4b}\includegraphics[scale=0.505]{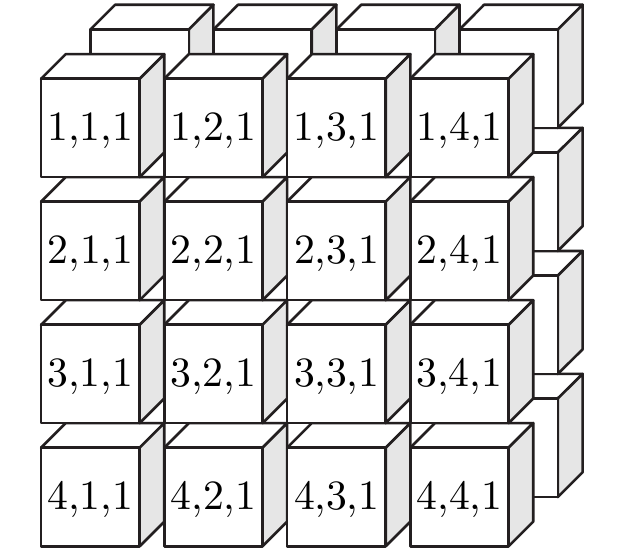}}
	\hfill		
	\vspace{-0.3cm}
  \caption{\label{fig:data}Example Tensor Representations.}
\end{figure}

\textbf{Local Tensor:} Our central data structure abstraction is a \texttt{TensorBlock} that represents a local tensor or a tile of a distributed tensor. Here, all single- and multi-threaded operations are implemented in Java for portability, but for compute-intensive operations we also support JNI calls to native BLAS libraries or custom C++ kernels. We provide two implementations of this \texttt{TensorBlock} abstraction:
\begin{itemize2}
\item \emph{\texttt{BasicTensorBlock} (Homogeneous):} A basic tensor is a linearized, multi-dimensional array of a single type (FP32, FP64, INT32, INT64, Bool, or String including JSON). We provide dense and sparse blocks and operations, which we apply based on the present sparsity.
\item \emph{\texttt{DataTensorBlock} (Heterogeneous):} A data tensor has a schema on the second dimension (see Figure~\ref{fig4a}), which generalizes 2D datasets. Internally, it is composed of multiple basic tensors for the given schema.
\end{itemize2}

\textbf{Distributed Tensors:} Our distributed tensor representation is a Spark RDD \cite{ZahariaCDDMMFSS12}---i.e., a distributed collection---of tensor indexes and fixed-size, independently-encoded blocks (\texttt{PairRDD<TensorIndexes,TensorBlock>}). Squared $1\text{K} \times 1\text{K}$ blocks in SystemML offer a good balance between amortized block overheads and moderate block sizes ($8\mb$ for dense), simplify join processing because blocks are always aligned, and allow local transformations for operations like transpose. However, fixed-size blocking for n-dimensional data---as shown in Figure~\ref{fig4b}---is challenging. We use a scheme of exponentially decreasing block sizes ($1024^2$, $128^3$, $32^4$, $16^5$, $8^6$, $8^7$), which similarly bounds the size to few megabytes and allows for local conversion. For example, on a 3D-tensor/matrix operation, we split each $1024^2$ matrix block into $64 \times 128^2$ blocks and perform the join, yielding again a 3D-tensor with $128^3$ blocking.

\textbf{Federated Tensors:} For federated operations we provide a federated tensor that is a metadata object holding references to---potentially remote---in-memory or distributed tensors. Subtensors cover disjoint index ranges of the tensor (most commonly subsets of rows or columns), and uncovered areas are zero. This representation is the basis for federated learning, as discussed in Section~\ref{sec:federated}, 
but also nicely generalizes operator placement to operations over multiple devices (e.g., 30\% of data on CPU, 70\% of data on GPU \cite{GowanlockKFW19}).

\section{Key Features and Directions}
\label{sec:dirs}

SystemDS shares several design aspects with other systems. In this section, we discuss some distinguishing features and research directions. However, we believe that building the overall system is of utmost importance for real impact and investigating these features in a realistic environment.

\subsection{Lineage and Reuse of Intermediates}
\label{sec:lineage}

\textbf{Efficient Lineage Tracing:} Exploratory data science has a high degree of redundancy and most frameworks lack model versioning and reproducibility. Hence, we provide built-in support for data provenance in terms of lineage tracing and exploitation. We see lineage as a key enabling technique for model versioning, reuse of intermediates, incremental maintenance, auto differentiation, more efficient buffer pool management, and debugging via query processing over lineage traces of different models or runs. In contrast to coarse-grained or data-oriented provenance, we focus on fine-grained lineage tracing of logical operations. We trace inputs (by name), literals, and all executed operations (including non-determinism like system-generated seeds) to maintain lineage DAGs of live variables. Additionally, for loops with few distinct control flow paths, we determine the lineage trace per path once, and track the taken path via a single lineage node for deduplication.

\textbf{Reuse of Intermediates:} Inspired by work on recycling intermediates in MonetDB \cite{IvanovaKNG09}, we then exploit this lineage for reusing redundantly computed intermediates, which are common in model selection workloads. We establish a cache, where intermediates are identified by their lineage (hash of the lineage DAG). Before executing an instruction, we update the output lineage and probe the cache for full or partial reuse. In contrast to existing work on coarse-grained reuse \cite{FernandezCWWMP18,LeeSCSWI18,SparksVKFR17,VartakTMZ18,XinMMLSP18,ZhangKR14}, partial reuse computes an output via a compensation plan over cached intermediates. For example, \texttt{steplm} in Example~\ref{ex:1} greedily adds features and performs what-if model training, which allows reusing intermediates from previous iterations, augmented by missing features.

\textbf{Status:} So far, we have integrated lineage tracing for local operations, lineage deduplication on while/for/parfor loops, basic caching and eviction policies, as well as full and partial reuse of intermediates. In the future, we intent to add rewrites for compiler-assisted reuse, more elaborate partial reuse, and query processing over collected lineage traces.

\subsection{Data Integration and Cleaning}
\label{sec:preparation}

\textbf{Semi-automated Data Preparation:} Fully-automated data integration, cleaning and preparation is rather unrealistic given its complexity. We aim to provide abstractions (e.g., data extraction, semantic type inference, schema alignment, entity linking, outlier and anomaly detection, missing value imputation, data augmentation, and feature transformations) that help a user compose data preparation pipelines. Providing support for efficient and accurate data preparation faces, however, many algorithmic and systems challenges. We start by adding respective built-in functions, where we aim for vectorized implementations to simplify inference and optimization as well as search space pruning via sparsity exploitation. For example, masking allows data slicing and missing value imputation (in chained-equation models \cite{mice}) via sequences of full matrix operations, which significantly simplifies the compilation into multi-threaded or distributed runtime plans. Overall, a key design choice is to retain the appearance of a stateless system by consuming pre-trained models and rules as tensors themselves.

\textbf{Efficient Data Ingestion:} Given these abstractions, efficient ingestion faces two more challenges. First, the number of external data formats is virtually unlimited and sometimes requires even custom parsers for nested data. Inspired by work on query processing over CSV, JSON, and binary data \cite{KarpathiotakisA16}, we aim to automatically generate code for efficient readers and writers from high-level descriptions of data formats. In this context, we further aim to avoid unnecessary parsing \cite{PalkarABZ18}, and unnecessary shuffling \cite{KunftKSRM17} by taking the entire preparation pipeline into account. Second, semi-automated data preparation  is still an exploratory process. Similar, to query processing over raw data \cite{AlagiannisBBIA12}, we aim to exploit the lineage-based reuse of intermediates and build dedicated access methods for linear algebra over raw data in multi-tenant data science workflows and federated ML.

\textbf{Status:} We already added built-in functions for schema detection, outlier detection, missing value imputation, data augmentation, model debugging, as well as additional input formats and feature transforms. In the future, we will incorporate state-of-the-art algorithms, improve accuracy for structured datasets, and focus on efficient data ingestion. 

\subsection{Federated ML}
\label{sec:federated}

\textbf{Motivation:} Early work on federated learning \cite{abs-1902-01046,McMahanMRHA17} shows great promise, but focuses on mini-batch ML algorithms over private data from mobile devices. We believe federated ML is broadly applicable in the enterprise as well, a view shared by recent work on enterprise model fusion \cite{VermaWM19}. First, it could create a spectrum of data ownership and sharing (private data, shared gradients/aggregates, shared data) enabling new markets and business models. Second, it could enable ML in geo-distributed or restricted environments, where data consolidation is infeasible.

\textbf{Federated ML Architecture:} Our basic design consists of multiple control programs, each having local data. A master control program holds the federated tensors (see Section~\ref{sec:data}) including connections to the remote workers waiting for commands. SystemDS then allows both, cross-data-center federation \cite{VulimiriCGKV15} (where each control program runs in a Spark cluster) as well as federation of individual endpoints. We aim to support linear algebra operations (and thus, all abstractions from Figure~\ref{fig:stack}), as well as distributed parameter servers over federated tensors. Special federated instructions process these federated tensors by pushing as much computation to the remote sites as possible, while complying with exchange constraints and leveraging means of cryptography\footnote{For example, homomorphic encryption \cite{Gentry09} allows multiply and add (and thus, matrix multiply) directly over encrypted data.}. We will further extend our existing parameter server to respect the boundaries of federated tensors as well.

\begin{example}[Federated MV Multiplication] \label{ex:2} For example, consider matrix-vector (MV) and vector-matrix (VM) multiplications on a federated matrix, where each site holds a partition of rows. For an MV multiplication, the master broadcasts the vector to the workers, lets them compute a local MV, collects the result vectors, and constructs the output vector via \texttt{rbind}. For a VM multiplication, the master sends only relevant vector slices to the workers, lets them compute a local VM, collects the result vectors, and computes the output vector by adding the individual results.
\end{example}

\textbf{Status:} So far, we have a basic integration of federated tensors and selected federated operations. In the future, we will focus on broad operation coverage, data preparation and cleaning of raw input data, efficient data exchange and materialization, as well as exchange constraints.

\subsection{Compiler and Runtime Improvements}

The stack of declarative abstractions from Figure~\ref{fig:stack} requires major extensions of the compiler and runtime. We are interested in the following related research directions:
\begin{itemize2}
\item \emph{ML \& Rules:} Complex ML apps often combine ML models and rules in meta models, which require dedicated compilation and verification techniques.
\item \emph{Size Propagation:} Propagating dimensions \cite{BoehmBERRSTT14} and sparsity \cite{Sommer0ERH19}---or at least lower and upper bounds---through control flow of the entire lifecycle is challenging but essential for cost-based optimization.
\item \emph{Operator Fusion \& Code Generation:} Fusion is a widely recognized optimization, but the potential for sparsity exploitation \cite{BoehmRHSEP18} is barely leveraged yet.
\item \emph{Lossless and Lossy Compression:} Recent work on lossless compression for linear algebra \cite{ElgoharyBHRR16,LiCZ00NP19} and quantization for DNN workloads need a systematic investigation regarding data tensors and federated ML. 
\item \emph{Cloud and Auto Scaling:} The stateless design and size inference also enable automatic resource optimization \cite{HuangBTRTR15} in cloud environments, which is still an obstacle.
\end{itemize2}

\section{Preliminary Experiments} 
\label{sec:exp}

Our experiments study the baseline performance of SystemDS and optimization opportunities across lifecycle tasks.

\subsection{Experimental Setting} 

We ran all experiments on a single node with two Intel Xeon E5-2620 CPUs\,@\,2.10-2.50\,GHz (24 virtual cores), $128\gb$ DDR3 RAM, and CentOS Linux 7.4. SystemDS\,0.1+ (as of 12/2019, v0.1 released 08/2019, forked from SystemML\,1.2 in 09/2018) uses OpenJDK 1.8.0 with $80\gb$ max and initial JVM heap sizes. The baselines are TensorFlow\,1.13.2 (07/2019), TensorFlow\,2.0.0 (10/2019), and Julia\,1.1.1 (05/2019). The workloads are (1) a hyper-parameter optimization script (HPO) that reads a CSV file, trains $k$ regression models with different regularization $\lambda$ (see \texttt{lmDS} in Figure~\ref{fig:steplm}), and stores the resulting models as a CSV file, and (2) a cross-validation script (CV) that reads a CSV file, runs $k$-fold cross-validation for \texttt{lmDS}, and stores all $k$ models as a CSV file. We generate synthetic dense and sparse data, use optimized TF and Julia scripts, and report the end-to-end runtime including I/O as the mean of 3 repetitions.

\subsection{Baseline Comparison}

For evaluating the baseline performance of SystemDS, we use a $100\text{K} \times 1\text{K}$ matrix $\mat{X}$ ($800\mb$ in-memory, $1.79\gb$ CSV file) and train $k \in (1,10,20,30,40,50,60,70)$ models. The main computation of \texttt{lmDS} is $\mat{X}^{\top}\mat{X}$ and $\mat{X}^{\top}\mat{y}$, which requires $100.2\gflop$ per model but is independent of $\lambda$. Figures~\ref{exp1a} and \ref{exp1b} compare TensorFlow\,1.3 with NumPy array (TF) and tensor (TF-G) outputs---where the latter constructs a single graph and thus, can eliminate common subexpressions---Julia, SystemDS (SysDS) and SystemDS with native Intel MKL BLAS library (SysDS-B). There are four main observations. First, multi-threaded I/O in SysDS yields better performance than TF or Julia for a single model because string-to-double parsing is compute-intensive. Second, our multi-threaded, cache-conscious Java matrix multiplications show good performance but are 2.1x slower than Julia's native operations because Java does not compile packed SIMD instructions. With native BLAS for dense matrix-matrix multiplication, SysDS-B then slightly outperforms Julia. For TF, we had to manually rewrite \texttt{tf.matmul(tf.matrix\_transpose(X), X)} into a fused API call to avoid excessive transpose costs. Third, Figure~\ref{exp1b} shows that SysDS largely outperforms Julia and TF on sparse  data (with sparsity=nnz/\#cells=0.1). TF has large transpose overhead as its sparse-dense matrix multiply lacks a fused call, while TF-G executes this transpose only once. Fourth, and most importantly, none of these systems is able to eliminate the redundant matrix multiplications.  

\subsection{Reuse of Intermediates}

\begin{figure}[!t]
	\centering
	\vspace{-0.1cm}
	\subfigure[Baselines Dense]{
	   \label{exp1a}\includegraphics[scale=0.39]{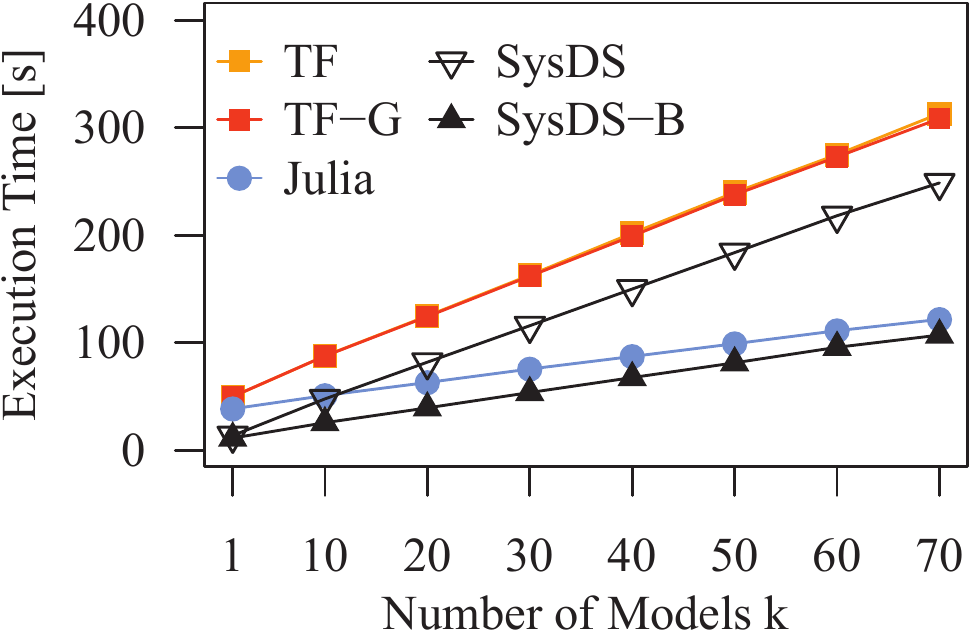}}
	\hfill	
	\subfigure[Baselines Sparse]{
	   \label{exp1b}\includegraphics[scale=0.39]{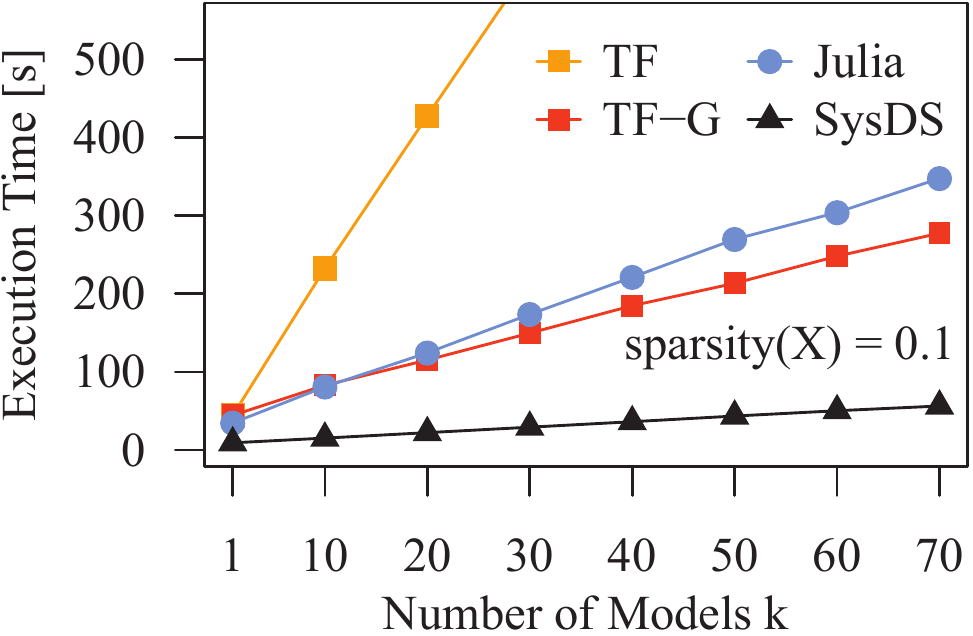}}~\vspace{-0.25cm}\\
	\subfigure[Reuse Dense]{
	   \label{exp2a}\includegraphics[scale=0.39]{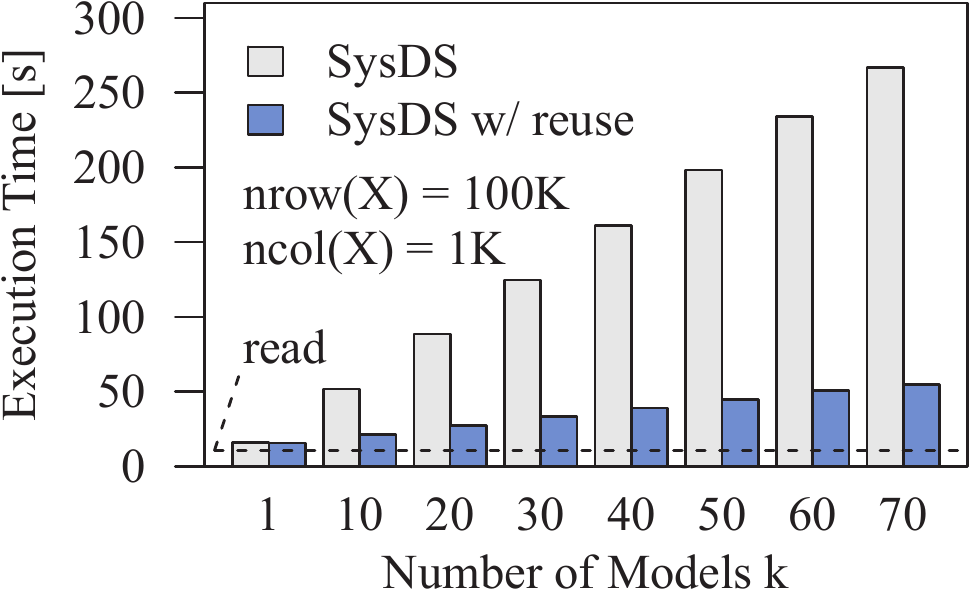}}
	\hfill	
	\subfigure[Reuse Sparse]{
	   \label{exp3}\includegraphics[scale=0.39]{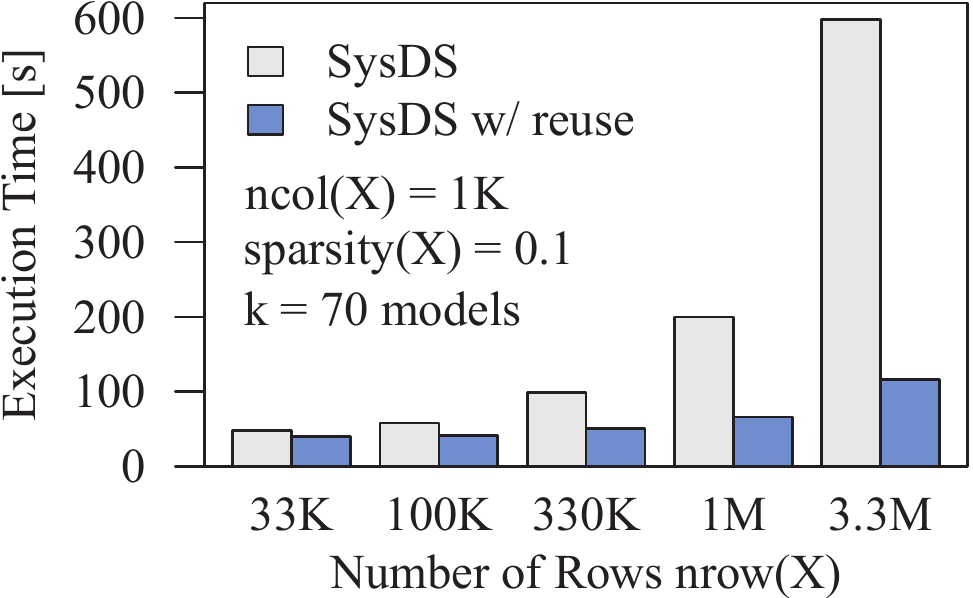}}
	\vspace{-0.4cm}
  \caption{\label{fig:exp}SystemDS Baseline Comparisons.}
\end{figure}

Figure~\ref{exp2a} shows the SystemDS results---for HPO over a $100\text{K} \times 1\text{K}$ dense input matrix $\mat{X}$---with enabled lineage-based reuse as described in Section~\ref{sec:lineage}. For one model, there is no redundancy. As the number of models increases, however, we see substantial improvements by reusing $\mat{X}^{\top}\mat{X}$ and $\mat{X}^{\top}\mat{y}$. Despite the I/O of 10.9s and several operations that are not subject to reuse, we get a 4.6x end-to-end speedup for 70 models. Figure~\ref{exp3} shows the impact of input data sizes by varying the number of rows in $\mat{X}$ (with sparsity=0.1). The larger the input, the higher the improvements because the remaining operations access only intermediates, whose size is independent of the number of rows. 

\begin{figure}[!b]
	\centering
	\subfigure[HPO Dense]{
	   \label{exp4a}\includegraphics[scale=0.39]{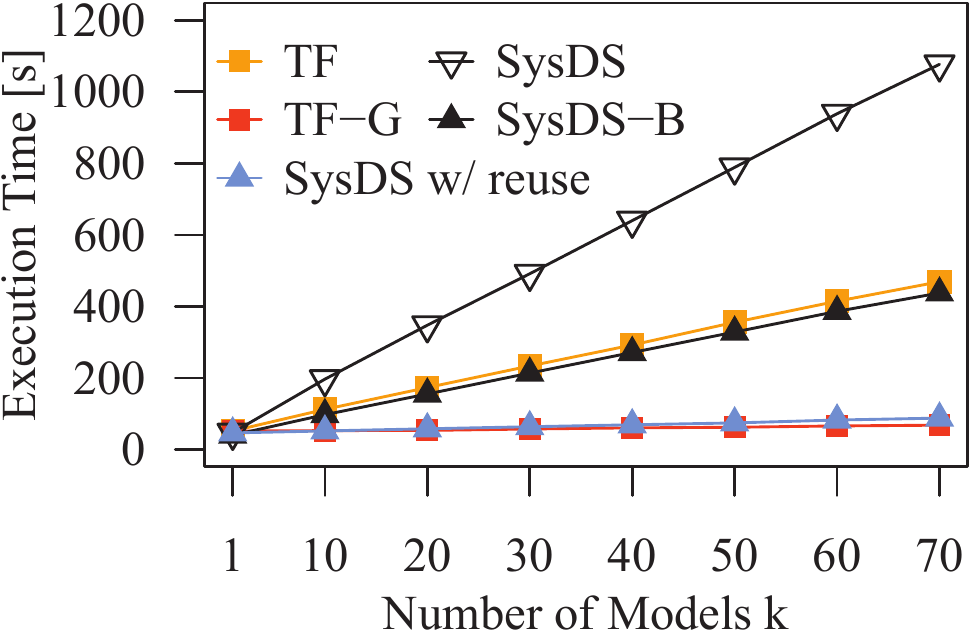}}
	\hfill	
	\subfigure[HPO Sparse]{
	   \label{exp4b}\includegraphics[scale=0.39]{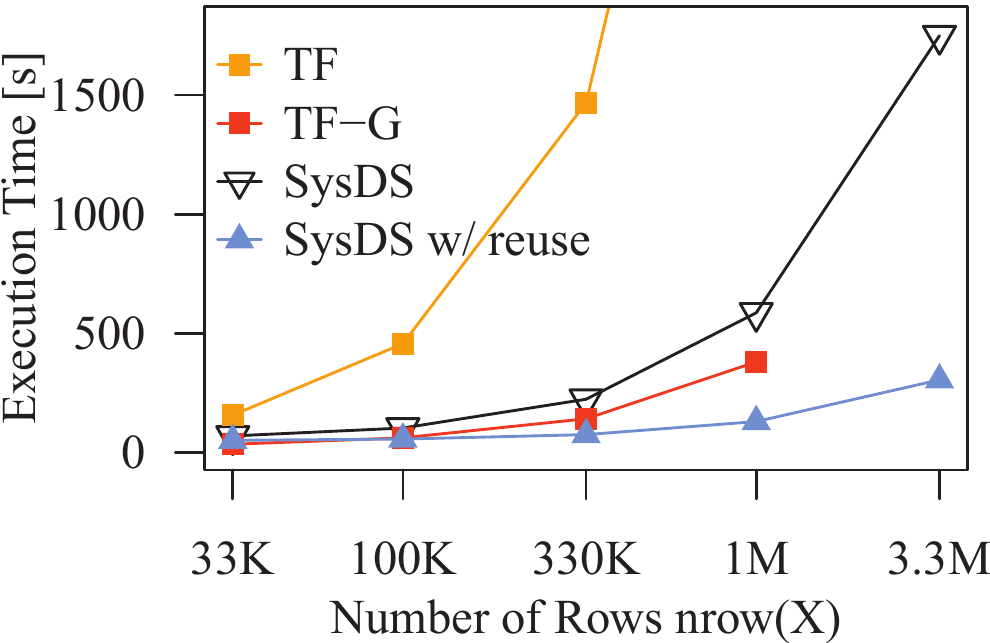}}
	\vspace{-0.4cm}
  \caption{\label{fig:exp4}SystemDS-TF2 Comparison (HPO).}
	\vspace{-0.1cm}
\end{figure}

\subsection{TensorFlow\,2 Comparison}

\textbf{HPO:} We further compare SystemDS with the recently released TensorFlow\,2.0 (compiled from sources), which introduced eager execution (TF) and lazy evaluation via AutoGraph \cite{autograph} (TF-G). Due to dependency conflicts, we ran these experiments in a VM with $80\gb$ RAM, 4 virtual cores, Ubuntu 18.04, and $70\gb$ max and initial JVM heap sizes. Figure~\ref{fig:exp4} shows the results for HPO, where TF-G is now able to reuse intermediates as well, while still causing substantial overhead for sparse data. Both TF and TF-G ran into segmentation faults for Figure~\ref{exp4b}, $3.3\text{M}$ though.

\begin{figure}[!t]
	\centering
	\vspace{-0.1cm}
	\subfigure[CV Dense]{
	   \label{exp5a}\includegraphics[scale=0.39]{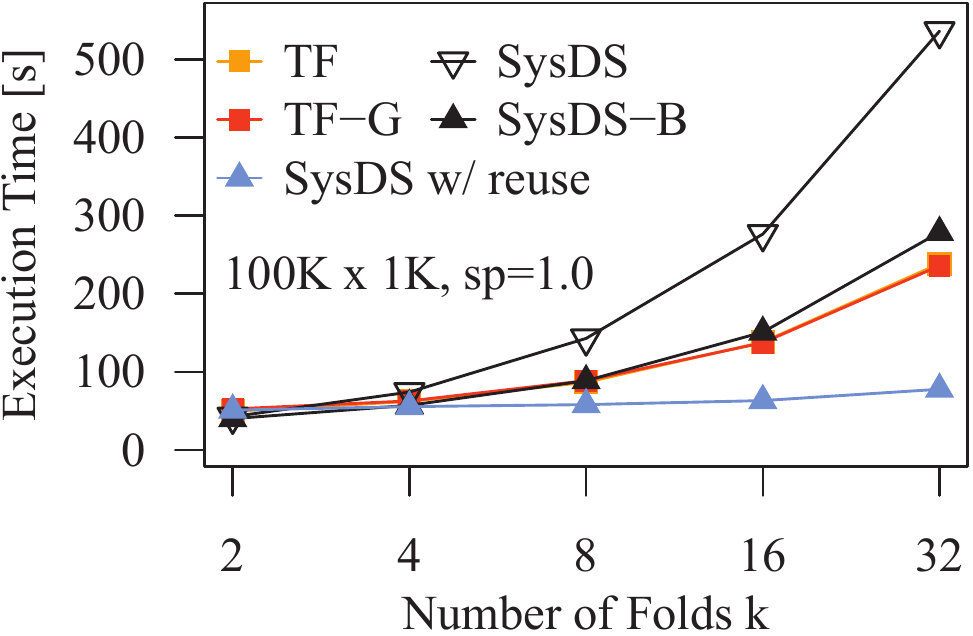}}
	\hfill	
	\subfigure[CV Sparse]{
	   \label{exp5b}\includegraphics[scale=0.39]{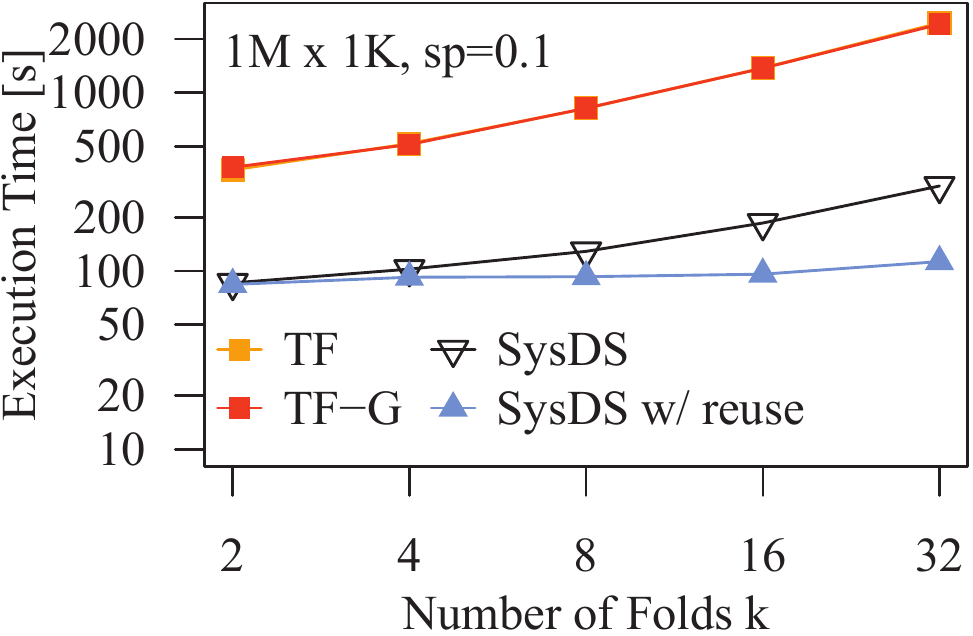}}
	\vspace{-0.4cm}
	\caption{\label{fig:exp5}SystemDS-TF2 Comparison (CV).}
	\vspace{-0.1cm}
\end{figure}

\textbf{CV:} Figure~\ref{fig:exp5} compares SystemDS 0.1+ and Tensor\-Flow 2.0 on the CV use case ($k$-fold cross validation, with leave- one-out). On dense data, TF/TF-G perform slightly better than SysDS-B, while on sparse data, TF/TF-G show again substantial overhead (log scale). Most importantly, only SysDS with reuse eliminates the fine-grained redundancy, which would be difficult in the AutoGraph model. Full reuse relies on rewriting \texttt{X=rbind(remove(foldsX,i))}, \texttt{y=rbind(remove(foldsY,i))}, and the matrix multiplications $\mat{X}^{\top}\mat{X}$ and $\mat{X}^{\top}\mat{y}$ during dynamic recompilation into multiplications of the individual folds (which are subject to reuse) and element-wise addition of these intermediates.

\section{Related Work} 

SystemDS has a broad focus and thus, there is lots of related work for individual aspects. Therefore, we focus on systems for data science lifecycle tasks and array processing.

\textbf{ML Systems for Data Science:} Several recent systems also aim to support the data science lifecycle in a scalable manner. First, Northstar \cite{Kraska18}---a collection of tools for interactive data science---includes Alpine Meadow \cite{ShangZBKECBUK19} for automatic feature pre-processing and AutoML based on existing ML libraries (e.g., Scikit-learn) and custom operators. Second, TensorFlow Extended (TFX) \cite{BaylorBCFFHHIJK17} provides components for data ingestion, validation, transformation, as well as model training, validation and serving. These components have different backends (e.g., transform in Apache Beam, train in TensorFlow) and are composed via orchestration tools like Apache Airflow or Kubeflow. Third, MLflow \cite{ZahariaCD0HKMNO18} provides means of model management (e.g., tracking experiments), project packaging, and model deployment. Alpine Meadow, TFX, and MLflow rely on existing ML libraries, while Weld \cite{PalkarTSSAZ17} focuses on a minimalistic but invasive, UDF-based IR for optimizing across different libraries. In contrast, SystemDS builds on a common DSL, provides its own compiler and runtime and thus, can exploit fine-grained optimization opportunities. Fourth, systems like AIDA \cite{DsilvaMK18} and Lara \cite{lara} aim at joint relational and linear algebra in data-science-centric specification languages that are mapped to a common IR and then executed via existing SQL engines, data-parallel frameworks, or numerical computing libraries. The design of SystemDS differentiates by support for tensors, distributed and federated linear algebra, and broad support for the end-to-end data science lifecyle.

\textbf{Array Processing:} Array databases (e.g., SciDB \cite{StonebrakerBPR11}) and array libraries (e.g., NumPy \cite{WaltCV11}, DL4J/NDArray) focus primarily on scientific computing and respective formats. While array databases require loading and schema design for efficient distributed operations, data scientist seem to favor stateless systems and functional R or Python libraries and DSLs. Scalability limitations are addressed by new Python libraries like dask \cite{dask} and xarray \cite{xarray} for distributed, multi-dimensional array processing. Although these libraries do not optimize for ML workloads, they are increasingly used by scikit-learn to provide distributed ML algorithms. In contrast, SystemDS aims to provide abstractions for a variety of data science lifecycle tasks and users, as well as efficient linear algebra and optimization across the lifecycle.

\section{Conclusions} 

To summarize, we described the vision and system architecture of SystemDS, an open-source ML system for end-to-end data science pipelines. Compared to SystemML, the major differences are (1) support for data science lifecycle tasks (e.g., data preparation, training, and debugging) and users with different expertise (ML researchers, data scientist, and domain experts), (2) support for local, distributed, and federated ML, and (3) the data model of heterogeneous data tensors. We also outlined selected research directions, and promising preliminary results. SystemDS is early work-in-progress, but throughout the next years (or decades), we will continuously improve it, leverage it in real-world applications, and use it for grounding our research in a real system. We encourage the DB and ML systems community to use SystemDS as a baseline or testbed for extensions. 

\section*{Acknowledgements} 

\small
We thank the entire Apache SystemML team for the initial code base of SystemDS, especially Shivakumar Vaithyanathan, Douglas R. Burdick, Michael Dusenberry, Deron Eriksson, Alexandre V. Evfimievski, Nakul Jindal, Faraz Makari Manshadi, Niketan Pansare, Frederick R. Reiss, Luciano Resende, Prithviraj Sen, Arvind Surve, Shirish Tatikonda, Yuanyuan Tian, Glenn Weidner, and many additional students, colleagues, and collaborators. Furthermore, we thank Svetlana Sagadeeva, Afan Secic, and Norbert Pfeifer for code contributions and additional use cases for SystemDS; our anonymous CIDR reviewers for their  detailed and thought-provoking comments; as well as Tilmann Rabl, Alireza Rezaei Mahdiraji, Sarah Osterburg, Steffen Zeuch, Volker Markl, and Claus Neubauer for detailed discussions in the ExDRa project. Several team members were funded through the first author's endowed professorship for data management (852799) and the program ``ICT of the Future'' -- both initiatives of the Austrian Ministry for Transport, Innovation, and Technology (BMVIT).

\small
\bibliographystyle{abbrv}
\bibliography{CIDR2020}

\begin{thebibliography}{10}

\bibitem{AbadiBCCDDDGIIK16}
M.~Abadi et~al.
\newblock {TensorFlow: A System for Large-Scale Machine Learning}.
\newblock In {\em {OSDI}}, 2016.

\bibitem{AlagiannisBBIA12}
I.~Alagiannis et~al.
\newblock {NoDB: Efficient Query Execution on Raw Data Files}.
\newblock In {\em {SIGMOD}}, 2012.

\bibitem{BaylorBCFFHHIJK17}
D.~Baylor et~al.
\newblock {TFX: A TensorFlow-Based Production- Scale Machine Learning
  Platform}.
\newblock In {\em {SIGKDD}}, 2017.

\bibitem{BernsteinM07}
P.~A. Bernstein and S.~Melnik.
\newblock {Model Management 2.0: Manipulating Richer Mappings}.
\newblock In {\em {SIGMOD}}, 2007.

\bibitem{BezansonEKS17}
J.~Bezanson et~al.
\newblock {Julia: A Fresh Approach to Numerical Computing}.
\newblock {\em {SIAM} Review}, 59(1), 2017.

\bibitem{BoehmTRSTBV14}
M.~Boehm et~al.
\newblock {Hybrid Parallelization Strategies for Large- Scale Machine Learning
  in SystemML}.
\newblock {\em {PVLDB}}, 7(7), 2014.

\bibitem{BoehmBERRSTT14}
M.~Boehm et~al.
\newblock {SystemML's Optimizer: Plan Generation for Large-Scale Machine
  Learning Programs}.
\newblock {\em {IEEE} Data Eng. Bull.}, 37(3), 2014.

\bibitem{BoehmDEEMPRRSST16}
M.~Boehm et~al.
\newblock {SystemML: Declarative Machine Learning on Spark}.
\newblock {\em {PVLDB}}, 9(13), 2016.

\bibitem{BoehmRHSEP18}
M.~Boehm et~al.
\newblock {On Optimizing Operator Fusion Plans for Large-Scale ML in SystemML}.
\newblock {\em {PVLDB}}, 11(12), 2018.

\bibitem{abs-1902-01046}
K.~Bonawitz et~al.
\newblock {Towards Federated Learning at Scale: System Design}.
\newblock In {\em {SysML}}, 2019.

\bibitem{CambroneroFSM17}
J.~Cambronero et~al.
\newblock {Query Optimization for Dynamic Imputation}.
\newblock {\em {PVLDB}}, 10(11), 2017.

\bibitem{ChandolaBK09}
V.~Chandola et~al.
\newblock {Anomaly Detection: A Survey}.
\newblock {\em {ACM} Comput. Surv.}, 41(3), 2009.

\bibitem{ChenLLLWWXXZZ15}
T.~Chen et~al.
\newblock {MXNet: A Flexible and Efficient Machine Learning Library for
  Heterogeneous Distributed Systems}.
\newblock {\em CoRR}, abs/1512.01274, 2015.

\bibitem{ChengQR12}
Y.~Cheng et~al.
\newblock {GLADE: Big Data Analytics Made Easy}.
\newblock In {\em {SIGMOD}}, 2012.

\bibitem{ChungKPTW19}
Y.~Chung et~al.
\newblock {Slice Finder: Automated Data Slicing for Model Validation}.
\newblock In {\em {ICDE}}, 2019.

\bibitem{CohenDDHW09}
J.~Cohen et~al.
\newblock {MAD Skills: New Analysis Practices for Big Data}.
\newblock {\em {PVLDB}}, 2(2), 2009.

\bibitem{autoaugment}
E.~D. Cubuk et~al.
\newblock {AutoAugment: Learning Augmentation Policies from Data}.
\newblock In {\em {CVPR}}, 2019.

\bibitem{dask}
Dask.
\newblock {\em Dask: Library for dynamic task scheduling}, 2016.

\bibitem{DeshpandeIR07}
A.~Deshpande et~al.
\newblock {Adaptive Query Processing}.
\newblock {\em Foundations and Trends in Databases}, 1(1), 2007.

\bibitem{DongR18}
X.~L. Dong and T.~Rekatsinas.
\newblock {Data Integration and Machine Learning: A Natural Synergy}.
\newblock In {\em {SIGMOD}}, 2018.

\bibitem{ElgoharyBHRR16}
A.~Elgohary et~al.
\newblock {Compressed Linear Algebra for Large-Scale Machine Learning}.
\newblock {\em {PVLDB}}, 9(12), 2016.

\bibitem{FengKRR12}
X.~Feng et~al.
\newblock {Towards a Unified Architecture for in-RDBMS Analytics}.
\newblock In {\em {SIGMOD}}, 2012.

\bibitem{FernandezCWWMP18}
R.~C. Fernandez et~al.
\newblock {Meta-Dataflows: Efficient Exploratory Dataflow Jobs}.
\newblock In {\em {SIGMOD}}, 2018.

\bibitem{FeurerKES0H19}
M.~Feurer et~al.
\newblock {Auto-sklearn: Efficient and Robust Automated Machine Learning}.
\newblock In {\em AML}. 2019.

\bibitem{GebalyAGKS14}
K.~E. Gebaly et~al.
\newblock {Interpretable and Informative Explanations of Outcomes}.
\newblock {\em {PVLDB}}, 8(1), 2014.

\bibitem{Gentry09}
C.~Gentry.
\newblock Fully homomorphic encryption using ideal lattices.
\newblock In {\em {STOC}}, 2009.

\bibitem{GhotingKPRSTTV11}
A.~Ghoting et~al.
\newblock {SystemML: Declarative Machine Learning on MapReduce}.
\newblock In {\em {ICDE}}, 2011.

\bibitem{GowanlockKFW19}
M.~Gowanlock et~al.
\newblock {Accelerating the Unacceleratable: Hybrid CPU/GPU Algorithms for
  Memory-Bound Database Primitives}.
\newblock In {\em DaMoN}, 2019.

\bibitem{HeerHK15}
J.~Heer et~al.
\newblock {Predictive Interaction for Data Transformation}.
\newblock In {\em {CIDR}}, 2015.

\bibitem{HeidariMIR19}
A.~Heidari et~al.
\newblock {HoloDetect: Few-Shot Learning for Error Detection}.
\newblock In {\em {SIGMOD}}, 2019.

\bibitem{HellersteinRSWFGNWFLK12}
J.~M. Hellerstein et~al.
\newblock {The MADlib Analytics Library or MAD Skills, the SQL}.
\newblock {\em {PVLDB}}, 5(12), 2012.

\bibitem{HodgeA04}
V.~J. Hodge and J.~Austin.
\newblock {A Survey of Outlier Detection Methodologies}.
\newblock {\em Artif. Intell. Rev.}, 22(2), 2004.

\bibitem{xarray}
S.~Hoyer and J.~Hamman.
\newblock {xarray: N-D labeled Arrays and Datasets in Python}.
\newblock {\em JORS}, 5(1), 2017.

\bibitem{HuangBTRTR15}
B.~Huang et~al.
\newblock {Resource Elasticity for Large-Scale Machine Learning}.
\newblock In {\em {SIGMOD}}, 2015.

\bibitem{HulsebosHBZSKDH19}
M.~Hulsebos et~al.
\newblock {Sherlock: A Deep Learning Approach to Semantic Data Type Detection}.
\newblock In {\em {KDD}}, 2019.

\bibitem{IvanovaKNG09}
M.~Ivanova et~al.
\newblock {An Architecture for Recycling Intermediates in a Column-store}.
\newblock In {\em {SIGMOD}}, 2009.

\bibitem{KandelPHH11}
S.~Kandel et~al.
\newblock {Wrangler: Interactive Visual Specification of Data Transformation
  Scripts}.
\newblock In {\em {CHI}}, 2011.

\bibitem{KarpathiotakisA16}
M.~Karpathiotakis et~al.
\newblock {Fast Queries Over Heterogeneous Data Through Engine Customization}.
\newblock {\em {PVLDB}}, 9(12), 2016.

\bibitem{KotthoffTHHL17}
L.~Kotthoff et~al.
\newblock {Auto-WEKA 2.0: Automatic model selection and hyperparameter
  optimization in WEKA}.
\newblock {\em JMLR}, 18, 2017.

\bibitem{Kraska18}
T.~Kraska.
\newblock {Northstar: An Interactive Data Science System}.
\newblock {\em {PVLDB}}, 11(12), 2018.

\bibitem{KumarNP15}
A.~Kumar et~al.
\newblock {Learning Generalized Linear Models Over Normalized Data}.
\newblock In {\em {SIGMOD}}, 2015.

\bibitem{KunftKSRM17}
A.~Kunft et~al.
\newblock {BlockJoin: Efficient Matrix Partitioning Through Joins}.
\newblock {\em {PVLDB}}, 10(13), 2017.

\bibitem{lara}
A.~Kunft et~al.
\newblock {An Intermediate Representation for Opti- mizing Machine Learning
  Pipelines}.
\newblock {\em {PVLDB}}, 12(11), 2019.

\bibitem{LeeSCSWI18}
Y.~Lee et~al.
\newblock {PRETZEL: Opening the Black Box of Machine Learning Prediction
  Serving Systems}.
\newblock In {\em {OSDI}}, 2018.

\bibitem{LiCZ00NP19}
F.~Li et~al.
\newblock {Tuple-oriented Compression for Large-scale Mini-batch SGD}.
\newblock In {\em {SIGMOD}}, 2019.

\bibitem{LuoGGPJ17}
S.~Luo et~al.
\newblock {Scalable Linear Algebra on a Relational Database System}.
\newblock In {\em {ICDE}}, 2017.

\bibitem{McMahanMRHA17}
B.~McMahan et~al.
\newblock {Communication-Efficient Learning of Deep Networks from Decentralized
  Data}.
\newblock In {\em {AISTATS}}, 2017.

\bibitem{MengBYSVLFTAOXX16}
X.~Meng et~al.
\newblock {MLlib: Machine Learning in Apache Spark}.
\newblock {\em JMLR}, 17, 2016.

\bibitem{autograph}
D.~Moldovan et~al.
\newblock {AutoGraph: Imperative-style Coding with Graph-based Performance}.
\newblock In {\em {SysML}}, 2019.

\bibitem{NikolicO18}
M.~Nikolic and D.~Olteanu.
\newblock {Incremental View Maintenance with Triple Lock Factorization
  Benefits}.
\newblock In {\em {SIGMOD}}, 2018.

\bibitem{OlsonM19}
R.~S. Olson and J.~H. Moore.
\newblock {TPOT: A Tree-Based Pipeline Optimization Tool for Automating Machine
  Learning}.
\newblock In {\em AML}. 2019.

\bibitem{PalkarTSSAZ17}
S.~Palkar et~al.
\newblock {A Common Runtime for High Performance Data Analysis}.
\newblock In {\em {CIDR}}, 2017.

\bibitem{PalkarABZ18}
S.~Palkar et~al.
\newblock {Filter Before You Parse: Faster Analytics on Raw Data with Sparser}.
\newblock {\em {PVLDB}}, 11(11), 2018.

\bibitem{PassingTHLSGK017}
L.~Passing et~al.
\newblock {SQL- and Operator-centric Data Analytics in Relational Main-Memory
  Databases}.
\newblock In {\em {EDBT}}, 2017.

\bibitem{pytorch}
A.~Paszke et~al.
\newblock {Automatic differentiation in PyTorch}.
\newblock 2017.

\bibitem{PyTorch2}
A.~Paszke et~al.
\newblock {PyTorch: An Imperative Style, High- Performance Deep Learning
  Library}.
\newblock In {\em {NeurIPS}}, 2019.

\bibitem{PedregosaVGMTGBPWDVPCBPD11}
F.~Pedregosa et~al.
\newblock {Scikit-learn: Machine Learning in Python}.
\newblock {\em JMLR}, 12, 2011.

\bibitem{PolyzotisRWZ18}
N.~Polyzotis et~al.
\newblock {Data Lifecycle Challenges in Production Machine Learning: A Survey}.
\newblock {\em {SIGMOD} Record}, 47(2), 2018.

\bibitem{RamanH01}
V.~Raman and J.~M. Hellerstein.
\newblock {Potter's Wheel: An Interactive Data Cleaning System}.
\newblock In {\em {VLDB}}, 2001.

\bibitem{RekatsinasCIR17}
T.~Rekatsinas et~al.
\newblock {HoloClean: Holistic Data Repairs with Probabilistic Inference}.
\newblock {\em {PVLDB}}, 10(11), 2017.

\bibitem{MahoutSamsara}
S.~Schelter et~al.
\newblock {Samsara: Declarative Machine Learning on Distributed Dataflow
  Systems}.
\newblock {\em {NIPS MLSys}}, 2016.

\bibitem{SchelterLSCBG18}
S.~Schelter et~al.
\newblock {Automating Large-Scale Data Quality Verification}.
\newblock {\em {PVLDB}}, 11(12), 2018.

\bibitem{SchleichOC16}
M.~Schleich et~al.
\newblock {Learning Linear Regression Models over Factorized Joins}.
\newblock In {\em {SIGMOD}}, 2016.

\bibitem{SculleyHGDPECYC15}
D.~Sculley et~al.
\newblock {Hidden Technical Debt in Machine Learning Systems}.
\newblock In {\em NIPS}, 2015.

\bibitem{ShangZBKECBUK19}
Z.~Shang et~al.
\newblock {Democratizing Data Science through Interactive Curation of ML
  Pipelines}.
\newblock In {\em {SIGMOD}}, 2019.

\bibitem{DsilvaMK18}
J.~V.~D. silva et~al.
\newblock {AIDA - Abstraction for Advanced In-Database Analytics}.
\newblock {\em {PVLDB}}, 11(11), 2018.

\bibitem{Sommer0ERH19}
J.~Sommer et~al.
\newblock {MNC: Structure-Exploiting Sparsity Estimation for Matrix
  Expressions}.
\newblock In {\em {SIGMOD}}, 2019.

\bibitem{SparksVKFR17}
E.~R. Sparks et~al.
\newblock {KeystoneML: Optimizing Pipelines for Large-Scale Advanced
  Analytics}.
\newblock In {\em {ICDE}}, 2017.

\bibitem{StonebrakerBPR11}
M.~Stonebraker et~al.
\newblock {The Architecture of SciDB}.
\newblock In {\em {SSDBM}}, 2011.

\bibitem{StonebrakerI18}
M.~Stonebraker and I.~F. Ilyas.
\newblock {Data Integration: The Current Status and the Way Forward}.
\newblock {\em {IEEE} Data Eng. Bull.}, 41(2), 2018.

\bibitem{mice}
S.~van Buuren and K.~Groothuis-Oudshoorn.
\newblock {mice: Multivariate Imputation by Chained Equations in R}.
\newblock {\em {J. of Stat. Software 2011}}, 20(1), 2011.

\bibitem{WaltCV11}
S.~van~der Walt et~al.
\newblock {The NumPy Array: A Structure for Efficient Numerical Computation}.
\newblock {\em Comp.S\&E}, 13(2), 2011.

\bibitem{VartakTMZ18}
M.~Vartak et~al.
\newblock {MISTIQUE: A System to Store and Query Model Intermediates for Model
  Diagnosis}.
\newblock In {\em {SIGMOD}}, 2018.

\bibitem{VenablesR02}
W.~N. Venables and B.~D. Ripley.
\newblock {\em Modern applied statistics with S, 4th Ed.}
\newblock Statistics and computing. Springer, 2002.

\bibitem{VermaWM19}
D.~C. Verma et~al.
\newblock {Federated AI for the Enterprise: A Web Services Based
  Implementation}.
\newblock In {\em {ICWS}}, 2019.

\bibitem{VulimiriCGKV15}
A.~Vulimiri et~al.
\newblock {WANalytics: Analytics for a Geo-Distributed Data-Intensive World}.
\newblock In {\em {CIDR}}, 2015.

\bibitem{XinMMLSP18}
D.~Xin et~al.
\newblock {Helix: Holistic Optimization for Accelerating Iterative Machine
  Learning}.
\newblock {\em {PVLDB}}, 12(4), 2018.

\bibitem{ZahariaCDDMMFSS12}
M.~Zaharia et~al.
\newblock {Resilient Distributed Datasets: A Fault-Tolerant Abstraction for
  In-Memory Cluster Computing}.
\newblock In {\em {NSDI}}, 2012.

\bibitem{ZahariaCD0HKMNO18}
M.~Zaharia et~al.
\newblock {Accelerating the Machine Learning Lifecycle with MLflow}.
\newblock {\em {IEEE} Data Eng. Bull.}, 41(4), 2018.

\bibitem{ZhangKR14}
C.~Zhang et~al.
\newblock {Materialization Optimizations for Feature Selection Workloads}.
\newblock In {\em {SIGMOD}}, 2014.

\bibitem{abs-1911-06311}
D.~Zhang et~al.
\newblock {Sato: Contextual Semantic Type Detection in Tables}.
\newblock {\em CoRR}, abs/1911.06311, 2019.

\end{thebibliography}
\enlargethispage{\baselineskip} 

\end{document}